\documentclass[twocolumn,showpacs,showkeywords,preprintnumbers,amsmath,amssymb]{revtex4}
\usepackage{graphicx}
\usepackage{dcolumn}
\usepackage{epsfig,graphicx,amsmath}
\begin{document}
\title{Approaches in evaluating two-time correlation function}

\author{Sintayehu Tesfa}
 \email{sint_tesfa@yahoo.com}
\affiliation{Department of Physics,
Dilla University, P. O. Box 419, Dilla, Ethiopia}
\date{\today}

 \begin{abstract}
Derivation of the procedures that can be applied in evaluating two-time correlation function in terms of coherent-state propagator and corresponding Q-function is presented. On the basis that the involved functions are generally exponential in nature, obtaining the two-time second-order correlation function is essentially claimed to be reduced to carrying out relatively simple integrations. Fundamentally, the time dependence of the operators is transferred to the density operator. Moreover, manipulation in reordering the operators is performed by applying the usual trace operation. With all details, it is basically observed that the two-time correlation can be readily determined once the pertinent coherent-state propagator or Q-function is known. Since working with c-number equation is far more handy than the associated operator equation, it is expected that the results derived in this contribution can aid in easing the otherwise involving mathematical rigor.
\end{abstract}

 \maketitle

\section{INTRODUCTION}

In many conceivable situations, a unique state determination through measurement is somewhat physically unattainable errand. It must be obvious that detection of  photon is one of such difficulty. In other words, photon measurement may not provide sufficient information about the associated state of the photon since the interaction of the radiation with the measuring device  is often complicated. Even in this context, experimental setups for determining the photon number in certain physical processes such as atomic excitation is fairly known despite  the compelling fact that the photon number is usually  very large. It is worth noting that the most common method of photon measurement relies on the photoelectric effect in which the photoelectrons librated by the photoionization are counted. It is also good to note that the associated detector is sensitive essentially to the  photoelectrons and fundamentally registers the current or the voltage induced by  these electrons. As a result, the involved devices essentially operate in absorptive mechanism since the measurement is destructive as the photon responsible for the  production of the photoelectron disappears. In addition to this, in the photon detection process, it is imperative assuming that each absorbed photon gives rise to no more than one electron and conversely each electron is librated only by one photon.

In actual experimental setup, it is worth noting that there can be different ways of measuring photons. One of the most obvious procedures in this respect is counting the number of photons produced by a single source with the aid of a single detector. Though this seems straightforward approach of measuring the photon, it essentially lacks a potential for revealing the correlation between the photons produced at different times and positions. Remarkably, it is a common knowledge nowadays, similar correlations are responsible for witnessing the nonclassical properties of the radiation. In relation to this, a number of experiments, basically, interferometric by nature have been performed where the photons delayed in time and space are counted \cite{n1781046}. Experimental arrangements, in which the photons delayed in time are counted, are generally designated as delayed coincidence measurements. If a light produced by a single source is orchestrated to travel over unequal distances like in the Michelson-Morley interferometer, an interference pattern is produced \cite{ajs34333}. Sometimes, the  photons separated in time and space can be counted by more than one detector. In such a case, studying the nature of the correlation between these photons might be required which makes the mechanism of evaluating the correlation between the two photons at different time (usually called two-time second-order correlation function) an integral part and parcel of optical measurements.

It is, therefore, imperative looking for various approaches of mathematically determining the two-time second-order correlation function in modern quantum optics. In light of this, the technique for calculating the two-time correlation function based on trajectory approach is developed in \cite{job679} whereas the derivation of its evolution equation is presented in \cite{jpp70251}. Moreover, the two-time correlation function of the harmonic oscillator is evaluated in \cite{pre562747}. The two-photon quantum correlation between Stockes and anti-Stockes radiation in the lambda three-level atomic system has also been studied using the master equation formulation and Onsager-Lax regression theorem \cite{pra72043811}. Furthermore, in recent years theoretical analysis of delayed coincidence of the cavity radiation of the two-level atom   has been addressed \cite{jmo56105,jmo551587,pra414083} under various context. Nonclassical features including photon anti-bunching and sub-Poisson photon statistics have been reported upon calculating the two-time second-order correlation function. There has been also a great deal of interest in studying the statistical and quantum properties of the radiation generated by different mechanisms in terms of various quantities including second-order \cite{jpsj692873} and multi-time correlations for open and closed systems \cite{pra75052108,jpb41225502}.

 In order to extend similar analysis to more involving situations, it is presumed to be advantageous introducing  alternative approaches for evaluating the correlations between photons arriving at the site of the detectors at separated times. Hence, primarily, the main objective of this contribution is directed to fill the gap in this respect. To achieve this goal, in the first place, the general concept of quantum correlation based on the seminal work of Glauber \cite{pr1302529} is introduced. In addition to this, the procedures of evaluating the two-time correlation function applying the most obvious Onsager-Lax theorem \cite{lou,pr1292342,pr37405}, the coherent-state propagator \cite{pra26451,pra291275} and the quasi-statistical distribution functions \cite{pr40749,zp461,prl10277,pr1312766} (particularly Q-function) are presented where the former is included for the sake of completeness. Taking the already available resources in this regard into consideration, it is anticipated that the later approaches can be of great help in calculating various correlations describable in the form of moments of the radiation at different times.

\section{Correlation Function: General remark}

Theory of photon detection requires complete description of the interaction of radiation with matter, although this consideration  is not taken into account in this contribution due to the involved complications and lack of complete knowledge of the interaction. In the process of photon measurement, since photon absorption mechanism is employed, the  detectors are presumed to be insensitive to the associated spontaneous emission. Consequently, the annihilation operator of the radiation field (${\bf\hat{E}^{(\dagger)}}$) is believed to be the one involved
in the counting process. In light of this, it is possible to propose that if the field undergoes a transition
from an initial state  ($\mid i\rangle$) to a final state
($\mid f\rangle$) in which a single photon has been absorbed, the
elements of the transition matrix can take the form
\begin{align}\label{5001}T_{if} = \langle f|\hat{E}^{(\dagger)}|i\rangle.\end{align}
Assuming the measuring device to be an ideal photon detector with frequency independent
absorption probability, it is not difficult to comprehend that the probability per
unit time at which a photon is absorbed at given position in space and time can be expressed as \cite{pr1302529}
\begin{align}\label{5002}W_{if} = \langle i\mid\hat{E}^{(-)}\mid f\rangle\langle f\mid
\hat{E}^{(\dagger)}\mid i\rangle,\end{align} where $({\bf\hat{E}^{(\dagger)}})^{\dagger} = {\bf\hat{E}^{(-)}}$.

In actual setting, the final state of the field would not be measured. But the measuring device registers the total count. In order to obtain the total count, it is necessary to
sum overall states of the field that can be reached at from the
initial state via absorption process. In light of this, the total counting
rate or average field intensity can be defined as
\begin{align}\label{5004}I({\bf{r}},t) = \sum_{f}\langle i\mid\hat{E}^{(-)}({\bf{r}},t)
\mid f\rangle\langle f\mid \hat{E}^{(\dagger)}({\bf{r}},t)\mid
i\rangle.\end{align} With the claim that the final states are complete, $\hat{I} = \sum_{f}|f\rangle\langle f|$, the average field
intensity can be rewritten as
\begin{align}\label{5006}I({\bf{r}},t) = \langle i\mid\hat{E}^{(-)}({\bf{r}},t)
\hat{E}^{(\dagger)}({\bf{r}},t)\mid i\rangle.\end{align} It is not difficult to observe that the expectation value in Eq. \eqref{5006} is quantifiable in view of the initial state alone. Moreover, it is noteworthy that
in the product of the field operators the creation operator
precedes the destruction operator which corresponds to the normal ordering.

In principle, it may appear easy and straightforward to conceive that recording
photon intensities using a single detector ensures exhaustive
measurement associated with the field \cite{prl622941,prl591903}. It, hence, turns out to be imperative looking for other possible mechanisms for carrying out reliable measurement on the field. In this respect, with the assumption that
there are two fields emerging from position ${\bf{r}}$ and
detected at separate times $t_{1}$ and $t_{2}$, the correlation
between the two photons can be quantified using the correlation
function defined by
\begin{align}\label{5007}G^{(1)}({\bf{r}};t_{1},t_{2}) = Tr(\hat{\rho}\hat{E}^{(-)}({{\bf{r}}},t_{1})
\hat{E}^{(\dagger)}({\bf{r}},t_{2})),\end{align} where $\hat{\rho}$ is the density operator that describes the state of the radiation field and $Tr$ is a shorthand for trace operation. Eq. \eqref{5007} stands for the
 two-time first-order correlation function and it is
usually found to be sufficient to account for the classical interference experiments.

It is advisable to resort back to statistical formulation since
 precise knowledge of the field is almost absent. It is  noticeable
that $\hat{\rho}$ corresponds to the initial state of the
radiation field. First and foremost, Eq. \eqref{5007} can be interpreted as the
transition probability for the detector atom while it absorbs a
photon from a field at position ${\bf{r}}$ in time between $t$ and
$t + dt$. In many instances, stationary fields are the common interest in
quantum optics whereby the correlation function of the field is
invariant under the displacement of the time variable. Hence the
correlation function $G^{(1)}({\bf{r}};t_{1},t_{2})$ is presumed to
depend only on $t_{1}$ and $t_{2}$ through their difference, that
is, $\tau = t_{2} - t_{1}$. On account of this consideration, it is possible to see that the
two-time first-order correlation function can be denoted as
$G^{(1)}({\bf{r}};t_{1},t_{2}) =
G^{(1)}({\bf{r}};\tau)$.

On the other hand, the joint probability for detecting one
photoionization at position ${\bf{r_{1}}}$ between $t_{1}$ and
$t_{1} + dt_{1}$ and another at ${\bf{r_{2}}}$ between $t_{2}$ and
$t_{2} + dt_{2}$ with $t_{1} < t_{2}$ is describable by the
two-time second-order correlation function defined by
\begin{widetext}
\begin{align}\label{5010}G^{(2)}({\bf{r}}_{1},{\bf{r}}_{2};t_{1},t_{2})
 = Tr(\hat{\rho}\hat{E}^{(-)}({\bf{r}}_{1},t_{1})\hat{E}^{(-)}({\bf{r}}_{2},t_{2})
\hat{E}^{(\dagger)}({\bf{r}}_{2},t_{2})
\hat{E}^{(\dagger)}({\bf{r}}_{1},t_{1})).\end{align}
\end{widetext}
It is not difficult to observe
that the right hand of Eq. \eqref{5010} is time ordered in which
the operators at earlier times come first and they are also normally ordered wherein creation operator comes first. Eq. \eqref{5010} generally defines the two-time second-order correlation often interpreted as
the photon delayed coincidences between the two photons.

The discussion up to now is based on the field operators. However, in many instances in quantum optics,
a normalized correlation function in terms of the radiation or boson operators may be required. In this line, the
first-order normalized correlation function with the application of the relation between field and boson operators turns out to be
\begin{align}\label{5011}g^{(1)}(\tau) =
\frac{\langle\hat{a}^{\dagger}(t)\hat{a}(t+\tau)\rangle}{
\langle\hat{a}^{\dagger}(t)\hat{a}(t)\rangle}.\end{align} In the same way, the second-order normalized two-time correlation function can be put in terms of creation and annihilation boson operators as
\begin{align}\label{5012}g^{(2)}(\tau) =
\frac{\langle\hat{a}^{\dagger}(t)\hat{a}^{\dagger}(t+\tau)\hat{a}(t+\tau)\hat{a}(t)\rangle}{
\langle\hat{a}^{\dagger}(t)\hat{a}(t)\rangle^{2}},\end{align}
where the radiation field is assumed to be statistically stationary.

It has been a subject of discussion in earlier communications that the normalized correlation
function is a vital tool in identifying the corresponding photon statistics. It is common knowledge that when the field satisfies the inequality,
\begin{align}\label{5013}g^{(2)}(\tau) < g^{(2)}(0),\end{align}
for all $\tau$ less than some critical time $\tau_{c}$, the
photon exhibits excess correlation for times less than $\tau_{c}$. This phenomenon represents photon
bunching as the photons tend to distribute themselves in bunches
rather than at random, since the correlation for photons arriving
at the same time ($\tau=0$) is greater than the ones coming at
separated time of $\tau$. This actually means that when such a light falls on the photon
detector more pairs of photons are detected closer together than
further apart.  The reverse of this situation corresponds to the phenomenon of  photon anti-bunching where fewer photon pairs are detected closer together. The phenomenon of photon anti-bunching,  in particular, is one of the possible ways by which the nonclassical features of the light is manifested. With the aid of this approach, photon anti-bunching is found to be a fundamental property that when a two-level atom interacts with the radiation in a view that a time needs to be elapsed whatever small may be before an atom absorbs the radiation after it successfully emits it \cite{jmo541759,jmo56105}. It is also possible to characterize the photon statistics of the
light via calculating the two-time second-order correlation function. In connection to this, it has been known for long that $g^{(2)}(\tau) = 1$
represents Poissonian,  $g^{(2)}(\tau) > 1$
super-Poissonian and $g^{(2)}(\tau) < 1$ sub-Poissonian  photon statistics.

\section{Procedures for evaluating two-time correlation function}

In practice, a solution of the density matrix
is not always sufficient to determine the two-time correlation
function. In many instances, it may be required to find the transition
probability distribution. In some cases, it is also possible to evaluate the two-time
correlation function employing the explicit form of a one-time
correlation function obtainable with the aid of the master equation
or the Langevin equation. Although  two-time second-order correlation function is very important in studying statistical and quantum properties of the radiation, its evaluation is not often straightforward and easy due to the time difference in the operators. Nevertheless, in the following, some alternative approaches are discussed in order to make the  calculation of the two-time second-order correlation more easy to handle.

\subsection{Onsager-Lax regression theorem}

In principle, the correlation function can  be readily derived if the time evolution of the corresponding operator is
known. This is, essentially, equivalent to the knowledge of the
solution of the Heisenberg or the quantum Langevin equation. But, in practice, finding the correlation of operators evaluated at two different times is not obvious as it might appeared from the outset. Quite often, in order to calculate the two-time expectation value, it is
desirable to make use of the utility offered by the Onsager-Lax theorem .  To this effect, the density operator at a time
$\tau$ with $\tau \geq 0$ is expressed in terms of the density
operator at earlier time $t = 0$ as
\begin{align}\label{5039}\hat{\rho}(\tau) = \hat{U}(\tau)\hat{\rho}(0)\hat{U}^{\dagger}(\tau),\end{align}
where $\hat{U}(\tau)$ is the usual evolution operator defined by
\begin{align}\label{5040}\hat{U}(\tau) = \exp(-i\hat{H}_{S}\times\tau),\end{align}
in which $\hat{H}_{S}$ is the system Hamiltonian.

Suppose the system under consideration is consistent with Makrovian approximation. This entails that the correlation between the system and reservoir at equal time is unimportant which implies that it is sufficient to write $\hat{\rho}_{SR}(t)=\hat{\rho}_{S}(t)\bigotimes\hat{\rho}_{R}(t)$. In this approximation, the evolution of a single-time expectation value can be expressed following the same reasoning as in Eq. \eqref{5039}  as
\begin{align}\label{tt}\langle\hat{A}(t+\tau)\rangle=Tr_{S}Tr_{R}(\hat{U}^{\dagger}(\tau)\hat{A}(t)\hat{U}(\tau)\hat{\rho}_{S}(t)
\bigotimes\hat{\rho}_{R}(t)).\end{align}
Using the cyclic property of trace operation, it is possible to see that
\begin{align}\label{tt}\langle\hat{A}(t+\tau)\rangle=Tr_{S}[\hat{A}(t)Tr_{R}(\hat{U}(\tau)\hat{\rho}_{S}(t)
\bigotimes\hat{\rho}_{R}(t)\hat{U}^{\dagger}(\tau))].\end{align}

Since $\hat{\rho}_{S}(t)\bigotimes\hat{\rho}_{R}(t)$ represents a single density operator ($\hat{\rho}_{SR}(t)$) that describes the combined system-reservoir, it is not difficult to note that
\begin{align}\label{tt}\langle\hat{A}(t+\tau)\rangle=Tr_{S}(\hat{A}(t)Tr_{R}(\hat{\rho}_{S}(t+\tau)
\bigotimes\hat{\rho}_{R}(t+\tau))),\end{align} which can also be rewritten as
\begin{align}\label{5041}\langle\hat{A}(t+\tau)\rangle\; = \sum_{j}G_{j}(\tau)\langle\hat{A}_{j}(t)\rangle,\end{align}
where $G_{j}(\tau)$'s are coefficients that depend on $\tau$. In the same manner, the two-time correlation function can be put in the form
\begin{align}\label{tt}\langle\hat{A}(t+\tau)\hat{B}(t)\rangle&=Tr_{S}Tr_{R}(\hat{U}^{\dagger}(\tau)\hat{A}(t)\hat{B}(t)\hat{U}(\tau)\notag\\&\times\hat{\rho}_{S}(t)
\bigotimes\hat{\rho}_{R}(t)).\end{align}
With the aid of the cyclic property of trace operation, one can write
\begin{align}\label{tt}\langle\hat{A}(t+\tau)\hat{B}(t)\rangle&=Tr_{S}(\hat{A}(t)\hat{B}(t)Tr_{R}(\hat{U}(\tau)\notag\\&\times\hat{\rho}_{S}(t)
\bigotimes\hat{\rho}_{R}(t)\hat{U}^{\dagger}(\tau))).\end{align}
Comparing with earlier discussion shows that
\begin{align}\label{5042}\langle\hat{A}(t+\tau)\hat{B}(t)\rangle\; = \sum_{j}G_{j}(\tau)\langle\hat{A}_{j}(t)\hat{B}_{j}(t)\rangle.\end{align}
 The procedure of transferring the time $\tau$ from the operator to a function $G(\tau)$ is described as Onsager-Lax or commonly known  as quantum regression theorem \cite{pr1292342,pr37405}. Fundamentally, what remains is obtaining $G(\tau)$ based on the
time evolution of the density operator that depends on the underlying physical system and accompanying existing circumstance.

\subsection{coherent-state propagator}

In the previous discussion, the time evolution of the quantum system to be studied is described by an operator $\hat{U}$  directly related to the pertinent system Hamiltonian according to Eq. \eqref{5040}. On account of the mathematical difficulty involved in manipulating the operators, it is found advantageous employing the corresponding  c-number equation. One of such formalisms is based on replacing the evolution operator with associated c-number function usually designated as coherent-state propagator \cite{pra26451,pra291275}. With the aid of this approach, theoretical investigation of the nonclassical features of the radiation generated by parametric oscillator \cite{oc151384} and spontaneously induced entanglement in the cavity radiation of $N$ two-level atoms \cite{jmo551683} have been reported recently. It has been observed that this consideration significantly simplifies the otherwise cumbersome process. It is, hence, expected based on earlier studies that using c-number representation can serve the purpose in easing the rigor of determining the two-time second-order correlation function. In this regard, Shao {\it{et al}}. \cite{jcp116507} have used path integral formulation to find two-time correlation function for systems connected with heat bath. With this motivation, in this section, the way of obtaining the  two-time second-order correlation function applying the coherent-state propagator would be developed. To
this end, an arbitrary function (correlation function) of the form
\begin{align}\label{5044}g(\tau) = \;\langle\hat{a}^{\dagger}(t +
\tau)\hat{a}(t)\rangle\end{align} is taken for clarity. It is straightforward to realize that this
expectation value can be expressed in the Heisenberg picture in
terms of the density operator at initial time ($\hat{\rho}(0)$) as
\begin{align}\label{5045}g(\tau) = Tr(\hat{\rho}(0)\hat{a}^{\dagger}(t +
\tau)\hat{a}(t)).\end{align} It would not be difficult to notice that the time dependence
can be transferred to the density operator using the fact
that $Tr(\hat{\rho}(0)\hat{A}(t)) = Tr(\hat{\rho}(t)\hat{A}(0))$
as
\begin{align}\label{5047}g(\tau) = Tr(\hat{\rho}(t)\hat{a}^{\dagger}(\tau)\hat{a}),\end{align}
where $\hat{\rho}(t)$ can readily be obtained from $\hat{\rho}(0)$
with the aid of Eq. \eqref{5039}, that is,
\begin{align}\label{5048}\hat{\rho}(t) =
\hat{U}(t)\hat{\rho}(0)\hat{U}^{\dagger}(t).\end{align}

By introducing the completeness relation for the coherent state,
$\hat{I} =\int\frac{d^{2}\alpha}{\pi}|\alpha\rangle\langle\alpha|$,
in Eq. \eqref{5047} along with the application of Eq. \eqref{5048}, it is possible to see
that
\begin{align}\label{5050}g(\tau) = \int\frac{d^{2}\alpha}{\pi}Tr(\hat{U}(t)\hat{\rho}(0)\hat{U}^{\dagger}(t)
\hat{a}^{\dagger}(\tau)\hat{a}|\alpha\rangle\langle\alpha|).\end{align}
On the basis of the cyclic property of trace operation, Eq. \eqref{5050} can be rewritten as
\begin{align}\label{5051}g(\tau) = \int\frac{d^{2}\alpha}{\pi}\alpha\langle\alpha|\hat{U}(t)\hat{\rho}(0)\hat{U}^{\dagger}(t)
\hat{a}^{\dagger}(\tau)|\alpha\rangle.\end{align}

Assuming the initial state of the system to be described by arbitrary state
$\mid\alpha_{0}\rangle$, Eq. \eqref{5051} can be put in the form
\begin{align}\label{5052}g(\tau) = \int\frac{d^{2}\alpha}{\pi}\alpha\langle\alpha|\hat{U}(t)
\mid\alpha_{0}\rangle\langle\alpha_{0}\mid\hat{U}^{\dagger}(t)
\hat{a}^{\dagger}(\tau)|\alpha\rangle.\end{align} Now upon defining, $K(\alpha,t|\beta,0) =
\;\langle\alpha|\hat{U}(t)|\beta\rangle$, as the
coherent-state propagator for a single-mode radiation, it is found
employing the completeness relation for coherent state once again that
\begin{align}\label{5054}\langle\alpha|\hat{U}(t)|\alpha_{0}\rangle\; =
\int\frac{d^{2}\alpha_{1}}{\pi}K(\alpha,t|\alpha_{1},0)\langle\alpha_{1}|\alpha_{0}\rangle.\end{align}
Furthermore, with the introduction of the completeness relation once again,
one can readily see that
\begin{align}\label{5055}\langle\alpha_{0}|\hat{U}^{\dagger}(t)\hat{a}^{\dagger}(\tau)|\alpha\rangle\; =
\int\frac{d^{2}\alpha_{2}}{\pi}\langle\alpha_{0}|\hat{U}^{\dagger}|\alpha_{2}\rangle
\langle\alpha_{2}|\hat{a}^{\dagger}(\tau)|\alpha\rangle,\end{align} which can also be rewritten following the same reasoning as
\begin{align}\label{5056}
\langle\alpha_{0}|\hat{U}^{\dagger}(t)\hat{a}^{\dagger}(\tau)|\alpha\rangle\;
&=
\int\frac{d^{2}\alpha_{2}}{\pi}\frac{d^{2}\alpha_{3}}{\pi}K^{*}(\alpha_{2},t|\alpha_{3},0)\notag\\&\times\langle\alpha_{0}|\alpha_{3}\rangle
\langle\alpha_{2}|\hat{a}^{\dagger}(\tau)|\alpha\rangle,\end{align}
where
$K^{*}(\alpha_{2},t\mid\alpha_{3},0)
=\langle\alpha_{3}\mid\hat{U}^{\dagger}(t)\mid\alpha_{2}\rangle$.

Following the same line of argument, it is not difficult to see that
\begin{align}\label{5058}
\langle\alpha_{2}|\hat{a}^{\dagger}(\tau)|\alpha\rangle\; =
Tr(\hat{\rho}'(0)\hat{a}^{\dagger}(\tau)),\end{align} where
 $\hat{\rho}'(0) =
 |\alpha\rangle\langle\alpha_{2}|$.
 Applying the property of trace operation, it is possible to shift the time
 dependence to this density operator as we have done before. That
 is,
\begin{align}\label{5060}
\langle\alpha_{2}|\hat{a}^{\dagger}(\tau)|\alpha\rangle\; =
Tr(\hat{\rho}'(\tau)\hat{a}^{\dagger}),\end{align} where
$\hat{\rho}'(\tau) =
\hat{U}(\tau)|\alpha\rangle\langle\alpha_{2}|\hat{U}^{\dagger}(\tau)$.
Inserting the completeness relation for a coherent state into Eq.
\eqref{5060} leads to
\begin{align}\label{5062}
\langle\alpha_{2}|\hat{a}^{\dagger}(\tau)|\alpha\rangle\; =
\int\frac{d^{2}\alpha_{4}}{\pi}\alpha^{*}_{4}\langle\alpha_{4}|\hat{\rho}'(\tau)|\alpha_{4}\rangle,\end{align} which can also be rewritten as
\begin{align}\label{5063}
\langle\alpha_{2}|\hat{a}^{\dagger}(\tau)|\alpha\rangle\; = \int
{d^{2}\alpha_{4}\over\pi}\alpha^{*}_{4}K(\alpha_{4},\tau;\alpha,0)K^{*}(\alpha_{4},\tau;\alpha_{2},0).
\end{align}
On account of Eqs. \eqref{5052}, \eqref{5055},  and
\eqref{5063}, one readily gets
\begin{widetext}
\begin{align}\label{5064}g(\tau) &= \int\frac{d^{2}\alpha}{\pi}\frac{d^{2}\alpha_{1}}{\pi}
\frac{d^{2}\alpha_{2}}{\pi}\frac{d^{2}\alpha_{3}}{\pi}{d^{2}\alpha_{4}\over\pi}
\alpha\alpha^{*}_{4}\langle\alpha_{1}|\alpha_{0}\rangle
\langle\alpha_{0}|\alpha_{3}\rangle
K(\alpha_{4},\tau|\alpha,\tau)K^{*}(\alpha_{4},\tau|\alpha_{2},0)
K(\alpha,t|\alpha_{1},0)K^{*}(\alpha_{2},t|\alpha_{3},0).\end{align}
\end{widetext}
It is not difficult to realize that the correlation function can be evaluated
using the coherent-state propagator in the same manner. What essentially remains to be done is to extrapolate this derivation to the case when there are four operators instead of two, adapt the coherent-state propagator for different variables, and then carry out the integration straightaway. It is worth noting that fundamentally similar results have been obtained in \cite{jcp116507}. Characteristically, the method of evaluating the coherent-state propagator for most general Hamiltonian is provided in Ref. \cite{pra465379}. It has been observed that the coherent-state propagator associated with the quadratic Hamiltonian can be expressed in exponential form, which makes the involved task quite simple despite the number of integrations to be performed. For instance, following the procedure introduced in \cite{pra465379}, the coherent-state propagator for N two-level atom in the cavity and free space \cite{thesis} and parametric oscillation \cite{oc151384} have been calculated and it is found to be represented by simple exponential function.

\subsection{Q-Function}

One of the methods applicable while c-number equation is used to study quantum properties instead of the pertinent operator equation is the quasi-statistical distributions. Quasi-statistical distributions are c-number functions related to the density operator in certain predetermined order. One of these functions is the Husimi Q-function which corresponds to the normal ordering of the density operator. Quite generally, Q-function can be employed in calculating various order of moments. In view of the earlier efforts, it is expected that making use of the advantageous offered by this function can ease the rigor of obtaining the two-time second-order correlation function. To this effect, it is noticeable that the Q-function that corresponds to a time-dependent density operator can be defined as
\begin{align}\label{5065}Q(\alpha) =
\frac{1}{\pi}\langle\alpha|\hat{U}(t)|\alpha_{0}\rangle\langle\alpha_{0}|\hat{U}^{\dagger}(t)|\alpha\rangle.\end{align}
Introducing the completeness relation for a coherent state twice shows that
\begin{align}\label{5066}Q(\alpha) &=
\frac{1}{\pi}\int\frac{d^{2}\alpha_{5}}{\pi}\frac{d^{2}\alpha_{6}}{\pi}\notag\\&\times
\langle\alpha|\hat{U}(\tau)|\alpha_{5}\rangle\langle\alpha_{5}|\alpha_{0}\rangle\langle\alpha_{0}|\alpha_{6}\rangle\langle\alpha_{6}|
\hat{U}^{\dagger}(t)|\alpha\rangle.\end{align} On the basis of the definition of the coherent-state propagator, one gets straightaway
\begin{align}\label{5067}Q(\alpha) &=
\frac{1}{\pi}\int\frac{d^{2}\alpha_{5}}{\pi}\frac{d^{2}\alpha_{6}}{\pi}\notag\\\times&
K(\alpha,\tau|\alpha_{5},0)K^{*}(\alpha,t|\alpha_{6},0)
\langle\alpha_{5}|\alpha_{0}\rangle\langle\alpha_{0}|\alpha_{6}\rangle.\end{align}
Taking this into account, it is possible to rewrite Eq. \eqref{5064} as
\begin{align}\label{5068}g(\tau) = \int\frac{d^{2}\alpha}{\pi}
d^{2}\alpha_{2}d^{2}\alpha_{4}
Q'(\alpha_{4},\alpha^{*}_{4},\tau)Q(\alpha,\alpha^{*}_{2},t)\alpha\alpha^{*}_{4}.\end{align} It is not difficult to realize that the Q-functions in Eq.
\eqref{5068} are the pertinent quasi-statical function representing the system described in terms of
different variables.

On the other hand, the time evolution of the quantum system can be directly obtained from the corresponding density operator. In this line, suppose a two-time correlation
function can be expressed as
\begin{align}\label{5070}g(\tau)=
Tr(\hat{a}^{\dagger}\hat{a}(\tau)\hat{\rho}(t)).\end{align} It is worth noting that
the density operator can be expanded in the normal order applying the power
series representation as
\begin{align}\label{5071}\hat{\rho}(t)=\sum_{l,m}C_{lm}(t)\hat{a}^{\dagger^{l}}\hat{a}^{m}.\end{align} Therefore,
 introducing the coherent state completeness relation
leads to
\begin{align}\label{5072}g(\tau)=
\int{d^{2}\alpha\over\pi}\sum_{l,m}C_{lm}(t)Tr(\hat{a}^{\dagger}\hat{a}(\tau)
\mid\alpha\rangle\langle\alpha\mid\hat{a}^{\dagger^{l}}\hat{a}^{m}).\end{align}
In view of the action of the boson operators on the coherent state,
$\langle\alpha\mid\hat{a}^{\dagger^{l}}=\alpha^{*^{l}}\langle\alpha\mid$ and
$\langle\alpha\mid\hat{a}^{m}=\left(\alpha+{\partial\over\partial\alpha^{*}}\right)^{m}
\langle\alpha\mid$, it is possible to see that
\begin{align}\label{5075}g(\tau)&=
\int{d^{2}\alpha\over\pi}\sum_{l,m}C_{lm}(t)\alpha^{*^{l}}
\left(\alpha+{\partial\over\partial\alpha^{*}}\right)^{m}\notag\\&\times Tr
(\hat{a}^{\dagger}\hat{a}(\tau)
\mid\alpha\rangle\langle\alpha\mid).\end{align} On the basis of
the fact that
\begin{align}\label{5076}Q(\alpha,\alpha^{*},t)={1\over\pi}\sum_{l,m}C_{lm}(t)\alpha^{*^{l}}\alpha^{m},\end{align}
while the operators are initially put in the normal order, one finds
\begin{align}\label{5077}g(\tau)=
\int d^{2}\alpha Q\left(\alpha^{*}, \alpha
+{\partial\over\partial\alpha^{*}},t\right)Tr
(\hat{a}^{\dagger}\hat{a}(\tau)
\mid\alpha\rangle\langle\alpha\mid).\end{align} Making use of the cyclic
permutation of trace operation results in
\begin{align}\label{5078}Tr(\hat{a}^{\dagger}\hat{a}(\tau)\mid\alpha\rangle\langle\alpha\mid)=\alpha^{*}
\langle\alpha\mid\hat{a}(\tau)\mid\alpha\rangle.\end{align}

Furthermore, it is possible to express
\begin{align}\label{5079}\langle\alpha\mid\hat{a}(\tau)\mid\alpha\rangle=Tr(\hat{a}(\tau)\hat{\rho}
),\end{align} where $\hat{\rho}=\mid\alpha\rangle\langle\alpha\mid$.
With no doubt, the time factor can be transferred to the density operator in light of earlier discussion. That is,
\begin{align}\label{5081}\langle\alpha\mid\hat{a}(\tau)\mid\alpha\rangle=Tr(\hat{a}\hat{\rho}(\tau)
).\end{align} Now upon introducing a coherent state completeness
relation once again, one can write
\begin{align}\label{5082}\langle\alpha\mid\hat{a}(\tau)\mid\alpha\rangle=
\int{d^{2}\beta\over\pi}\beta
Tr(\mid\beta\rangle\langle\beta\mid\hat{\rho}(\tau)).\end{align}
With the aid of the definition of the Q-function in terms of the density
operator, one can see that
\begin{align}\label{5083}\langle\alpha\mid\hat{a}(\tau)\mid\alpha\rangle=\int
d^{2}\beta Q(\beta,\beta^{*},\tau)\beta.\end{align} Hence on
account of Eqs. \eqref{5077} and \eqref{5083}, one finally obtains
\begin{align}\label{5084}g(\tau)=\int
d^{2}\alpha d^{2}\beta Q\left(\alpha^{*},
\alpha+{\partial\over\partial\alpha^{*}},t\right)Q(\beta^{*},\beta,\tau)\alpha^{*}\beta.\end{align}

The Q-functions in Eq. \eqref{5084}  are also the same
Q-function pertinent to the quantum system under consideration in terms
of different variables. In the same way this approach can be extrapolated when there are more than two operators.

\section{Conclusion}

Detailed derivation of various approaches with which the two-time correlation function can be evaluated is presented. It is assumed that employing c-number equations instead of the corresponding operator equations eases the involved mathematical rigor. It is basically shown that the two-time second-order correlation function can be obtained from the pertinent coherent-state propagator and Q-function. Since the operators, consequently the photons, are presumed to be described at different times in present contribution, the coherent-state propagator and Q-function representing the quantum system under consideration should be defined in terms of specially different time parameters. This entails that the two-time second-order correlation function is expressed in terms of these functions that can be associated with different alternatives. In the view that quite significant number of quantum systems have quadratic Hamiltonian, the corresponding coherent-state propagator and Q-function are claimed to be well behaved exponential functions. Therefore, though the number of integrations to be carried out are found to be large undoubtedly they reduce to quite ordinary standard integrals. The possibility of rewriting the approaches following from applying the coherent-state propagator in terms of the associated Q-function is believed to be essential in verifying the obtained results. In the same way, the possibility of rewriting the way of obtaining the two-time second-order correlation making use of Q-function in terms of different variables can also provide an alternative means of determining it. It is hence expected that the detailed derivation presented in this work lays a foundation for viable approach of obtaining correlations of various moments evaluated at two different times. With no doubt, this procedure can readily be employed in evaluating various quantum correlations at equal time.


\begin{thebibliography}{1}
\bibitem{n1781046}  Brown R and  Twiss R Q 1956 {\it{Nature}} 1956 {\bf{178}} 1046
\bibitem{ajs34333}  Michelson A A and Morely E W 1887 {\it{Amer. J. Sci.}} {\bf{34}} 333
\bibitem{job679}  Karpati A,  Adam P and  Janszky J 2004 {\it{J. Opt. B: Quantum. Semiclass. Opt.}} {\bf{6}} 79
\bibitem{jpp70251}  Erofeey V I 2004 {\it{J. Plas. Phys.}} {\bf{70}} 251
\bibitem{pre562747}  Okumura K and  Tanimura Y 1997 {\it{Phys. Rev. E}} {\bf{56}} 2747
\bibitem{pra72043811}  Patnaik A K,  Agarwal G S,  Oai C H R and  Scully M O 2005 {\it{Phys. Rev. A}} {\bf{72}} 043811
\bibitem{jmo56105}  Tesfa S 2009 {\it{J. Mod. Opt.}} {\bf{56}} 105
\bibitem{pra414083}  D'Souza R,  Jayaroa A S and  Lawande S V 1990 {\it{Phys. Rev. A}} {\bf{41}} 4083
\bibitem{jmo551587}  Tesfa S 2008 {\it{J. Mod. Opt.}} {\bf{55}} 1587
\bibitem{jpsj692873}  Feng X L,  Zhang J T and  Xu Z Z 2000 {\it{J. Phys. Soc. Jpn}} {\bf{69}} 2873
\bibitem{jpb41225502}  Das S and  Agarwal G S 2008 {\it{J. Phys. B: At. Mol. Opt. Phys.}} {\bf{41}} 225501
\bibitem{pra75052108}  Alonso D and  de Vega I 2007 {\it{Phys. Rev. A}} {\bf{75}} 052108
\bibitem{pr1302529}  Glauber R J 1963 {\it{Phys. Rev.}} {\bf{130}} 2529
\bibitem{lou}  Louisell W H 1973 {\it{Quantum Statistical Properties of Radiation}} (Wiley, New York)
\bibitem{pr1292342}  Lax M 1963 {\it{Phys. Rev.}} {\bf{129}} 2342;  Lax M 1967 {\it{Phys. Rev.}} {\bf{157}} 213;  Lax M 2000 {\it{Opt.Commun.}} {\bf{179}} 463
\bibitem{pr37405}  Onsager L 1931 {\it{Phys. Rev.}} {\bf{37}} 405
\bibitem{pra291275}  Hillery M and  Zubairy M S 1984 {\it{Phys. Rev. A}} {\bf{29}} 1275
\bibitem{pra26451}  Hillery M and  Zubairy M S 1982 {\it{Phys. Rev. A}} {\bf{26}} 451
\bibitem{pr40749}  Winger E P 1932 {\it{Phys. Rev.}} {\bf{40}} 749
\bibitem{zp461}  Weyl H 1927 {\it{Z. Phys.}} {\bf{46}} 1
\bibitem{prl10277}  Sudarshan E C G 1963 {\it{Phys. Rev. Lett.}} {\bf{10}} 277
\bibitem{pr1312766}  Glauber R J 1963 {\it{Phys. Rev.}} {\bf{131}} 2766
\bibitem{prl622941}  Ou Z Y and  Mandel L 1989 {\it{Phys. Rev. Lett.}} {\bf{62}} 2941
\bibitem{prl591903}  Ghosh R and  Mandel L 1987 {\it{Phys. Rev. Lett.}} {\bf{59}} 1903
\bibitem{jmo541759}  Tesfa S 2007 {\it{J. Mod. Opt.}} {\bf{54}} 1759
\bibitem{oc151384}  Daniel B and  Fesseha K 1998 {\it{Opt. Commun.}} {\bf{151}} 384
\bibitem{jmo551683}  Tesfa S 2008 {\it{J. Mod. Opt.}} {\bf{55}} 1683
\bibitem{jcp116507}  Shao J and  Marki N 2002 {\it{J. Chem. Phys.}} {\bf{116}} 507
\bibitem{pra465379}  Fesseha K 1992 {\it{Phys. Rev. A}} {\bf{46}} 5379
\bibitem{thesis}  Tesfa S 1997 (MSc Thesis, Addis Ababa University)


\end{thebibliography}
\end{document}